\documentclass[a4paper]{jpconf}
\usepackage{graphicx}
\usepackage{times,latexsym,euscript,amstext,amssymb,amsbsy,amsmath,amsfonts,fancyhdr}

\newcommand{\ket}{\rangle}

\begin{document}
\title{Electromagnetic reactions from coupled-cluster theory}

\author{Sonia Bacca$^{1,2,3}$, Mirko Miorelli$^{2,4}$, Gaute Hagen$^{5,6}$}

\address{$^1$ Institut f\"ur Kernphysik, Johannes Gutenberg-Universit\"at Mainz, 55128 Mainz, Germany}
\address{$^2$ TRIUMF, 4004 Wesbrook Mall, Vancouver, BC V6T 2A3, Canada}
\address{$^3$ Department of Physics and Astronomy, University of Manitoba, Winnipeg, MB, Canada R3T 2N2}
\address{$^4$ Department of Physics and Astronomy, University of British Columbia, Vancouver, BC, V6T 1Z4, Canada}
\address{$^5$ Physics Division, Oak Ridge National Laboratory, Oak Ridge, TN 37831, USA}
\address{$^6$ Department of Physics and Astronomy, University of Tennessee, Knoxville, TN 37996,
  USA}

\ead{s.bacca@uni-mainz.de, miorelli@triumf.ca, hageng@ornl.gov}

\begin{abstract}
We  review recent results  for  electromagnetic reactions and related  sum rules in light and medium-mass nuclei obtained from  coupled-cluster theory. In particular, we highlight our recent computations of the photodisintegration cross section of $^{40}$Ca and of the electric dipole polarizability for oxygen and calcium isotopes. We also provide new results for the Coulomb sum rule for $^4$He and $^{16}$O. For $^4$He  we perform a thorough comparison of coupled-cluster theory with exact hyperspherical harmonics.
\end{abstract}

\section{Introduction}
Electromagnetic reactions have traditionally been a golden tool for studying nuclear dynamics. They lead, for instance, to the discovery of giant dipole resonances and their interpretation in terms of collective modes~\cite{GoT48,steinwedel1950}.  
While a complete body of data has been collected over the years for stable nuclei, only recently have we begun to elucidate them in terms of first principle calculations. This was achieved thanks to the introduction of a new  ab initio computational tool, obtained from a coupled cluster theory formulation of the Lorentz integral transform method~\cite{Bacca:2013dma}, which we will review below.

The key ingredient to study reactions induced by external electromagnetic probes, such as photons or electrons, is the nuclear response function. The latter is defined as
\begin{equation} 
R(\omega,q)=\sum_f \left|\langle \psi_f | \hat{\Theta}(q)|\psi_0\rangle \right|^2\delta\left(E_f-E_0-\omega \right),
\label{resp}
\end{equation}
where $\hat{\Theta}(q)$ is the excitation operator, which will in general depend on the momentum-transfer $q$ and whose form will be specific to the considered probe. The nuclear response function is a dynamical observables which requires knowledge on the whole spectrum of the nucleus, being $|\psi_0\rangle $ and $|\psi_f\rangle$ the ground- and excited-states, respectively.
The difficulty of studying Eq.~(\ref{resp}) lies in the fact that one needs, in principle, to 
calculate all the excited states $|\psi_f\rangle $, including all open channels in the continuum. While this is possible for two- and three-body problems, starting at four nucleons it can only be done with restrictions in energy and/or with approximations.  To overcome this complication, one can employ the Lorentz integral transform (LIT)~\cite{Efl07} and reduce the problem to the solution of a bound state equation. In this approach, rather than  directly calculating $R(\omega, q)$, one computes the following transform
\begin{equation} \label{lorenzo}
  {L}(q,\omega_0,\Gamma )=\int_{\omega_{\rm th}}^{\infty} d\omega \frac{R(\omega,q)}{(\omega -\omega_0)
               ^2+\Gamma^2}\:\mbox{.}
\end{equation}   
Here $\omega_{\rm th}$ is the threshold energy and $\Gamma > 0 $~is the Lorentzian width, 
which plays the role of a resolution parameter. By
using the closure relation one finds
\begin{equation}
\label{lorenzog} { L}(q,z)= \langle \psi_0 |
{\hat{\Theta}}^{\dagger}(q)\frac{1}{\hat
  {H}-z^*}\frac{1}{\hat{H}-z}\hat{\Theta}(q)|\psi_0 \rangle = \langle
\widetilde{\psi} | \widetilde{\psi} \rangle \:\mbox{,} 
\end{equation}
where we introduced the complex energy
$z=E_0+\omega_0+i\Gamma$ and $\hat{H}$ is the nuclear Hamiltonian, which will be taken from chiral effective field theory~\cite{Epelbaum09,Machleidt11}.
The LIT of the response function, $ {L}(q,z)$,
can be computed directly by solving the Schr{\"o}dinger-like equation
\begin{equation} \label{psi1}
  (\hat{H}-z)|\widetilde{\psi}\rangle =\hat
   {\Theta}(q) | \psi_0\rangle \: \hspace{1cm} 
\end{equation}
for different values of $z$. The solution
$|\widetilde{\psi}\rangle$ has bound-state-like asymptotic,  thus  $L(q,z)$ can be calculated
even for $A>3$ with any good bound-state method, such as hyperspherical harmonics expansions~\cite{barnea2001}.
 Results for
 $R(q,\omega)$ are  obtained from an
 inversion of the
LIT and are independent
on the choice of $\Gamma$~\cite{Efl07}.

Recently, we have merged the advantage
of the LIT method with the mild computational
scaling that characterizes coupled-cluster theory~\cite{hagen2013c} for increasing mass number,
obtaining a new computational tool~\cite{Bacca:2013dma}.  
In coupled-cluster theory, as originally introduced by Coester and K{\"u}mmel~\cite{coester1960}, the exact many-body wave function is written in the exponential {\it ansatz} 
\begin{equation}|\Psi_0\rangle=\exp{(T)}|{\rm \Phi_0}\rangle\,,
\end{equation}
 where $|{\rm \Phi_0}\rangle$ is a Slater determinant. The operator $T$, typically expanded in $n$-particle-$n$-hole excitations (or clusters) $T=T_1+T_2+\dots+T_A$, is responsible for introducing correlations.

The LIT method in the coupled-cluster language requires the solution of a bound-state equation
\begin{equation} \label{calc_rr}
 (\overline{H}-z) 
 |{\tilde{\Psi}}_R(z)\ket = \overline{\Theta}(q) | 0_R \ket\;,
\end{equation}
where the  right ground-state $ | 0_R \ket\equiv|  \Phi_0 \ket $, while $\overline{H}$ and $\overline{\Theta}$ are the similarity transformed Hamiltonian and excitation operators, respectively
\begin{eqnarray}\label{hbar}
 \overline{H} &=& \exp(-T) \hat H \exp(T)\, ,\\
 \overline{\Theta}& =& \exp(-T) \hat \Theta \exp(T) \,.
\end{eqnarray}
The solution of Eq.~(\ref{calc_rr}) is found as linear superposition of particle-hole excitation on top of the reference Slater determinant as
\begin{eqnarray} \label{eom1}\nonumber
|\widetilde{\Psi}_R(z)\ket &=&{\cal R}(z) | \Phi_0\ket = ({\mathcal R}_1+{\mathcal R}_2 + \dots + {\mathcal R}_A)|\Phi_0\ket \;.
\end{eqnarray}

 In Ref.~\cite{Bacca:2013dma} we have adopted the most commonly used approximation, i.e., the single and doubles scheme (CCSD), where $T=T_1+T_2$ and ${\mathcal R}={\mathcal R}_1+{\mathcal R}_2$ and have 
 successfully benchmarked this method  with exact hyperspherical harmonics on $^4$He, showing that the CCSD approximation differs only by 1-2$\%$ from the exact result.

 Because the computational cost of the coupled-cluster method scales mildly with respect to
the mass number, this theory  paves the way for many future investigations of electroweak reactions in medium-mass nuclei, as it provides an {\it ab initio} approach where the continuum is properly taken into account.

\section{The photodisintegration cross section}

The first observables where we used the power of coupled-cluster theory to address medium-mass nuclei has been the  photodisintegration cross section.  In the unretarded dipole approximation, the photodisintegration cross section is 
\begin{equation} 
\sigma_{\gamma}(\omega)=4\pi^2 \alpha \omega R(\omega)\,,
\label{cs}
\end{equation}
where  $R(\omega)$ is the dipole response function, with $\omega=q$ for real photons.
The excitation operator is  the translationally invariant dipole operator
\begin{equation}
\label{dip}
\hat{ \Theta}=\sum_k^A \left({\bf r}_k - {\bf R}_{\rm cm}  \right) \left( \frac{1+\tau^3_k}{2} \right)\,,
\end{equation}
where ${\bf r}_k$ and ${\bf R}_{\rm cm}$ are the coordinates of the $k$-th particle and
the center-of-mass, respectively, while $(1+\tau^3_k)/2$  defines
the projection operator on the $Z$ protons.

\begin{figure}
\begin{center}
\includegraphics*[width=9cm]{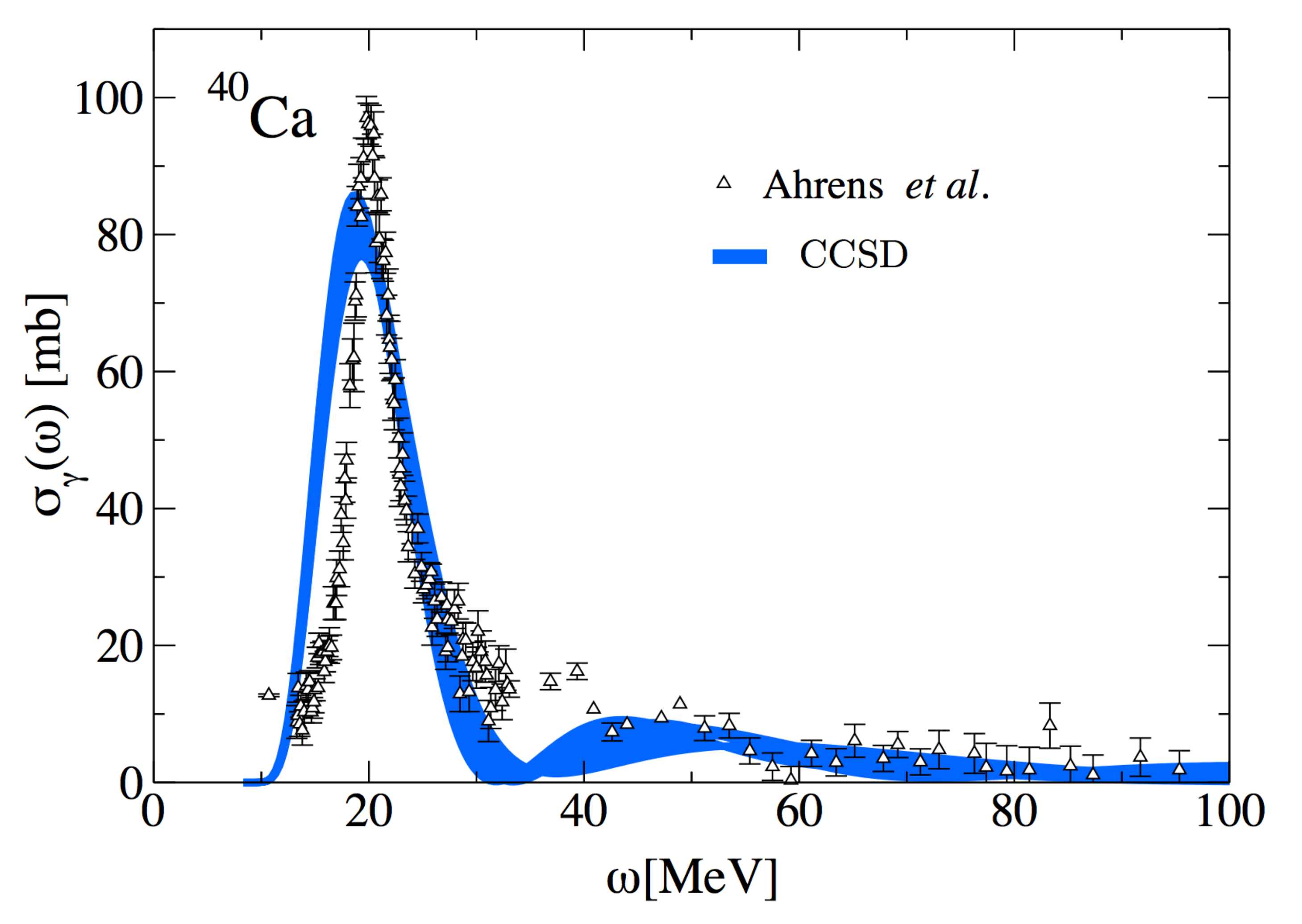}
\end{center}
\caption{\label{fig_Ca40}$^{40}$Ca photodisintegration cross section at the CCSD level~\cite{Mirko} compared to experimental data by Ahrens {\it et al.}~\cite{ahrens1975}. Result are obtained with a two-nucleon chiral force~\cite{Entem03}. Figure adapted from Ref.~\cite{Mirko}.}
\end{figure}

By solving Eq.~(\ref{calc_rr}) with the dipole operator in Eq.~(\ref{dip}) we have been able, e.g., to calculate
  the photodisintegration cross section of $^{40}$Ca~\cite{Mirko} using a realistic two-body chiral potential~\cite{Entem03}. Results are shown in Fig.~\ref{fig_Ca40}. The width of the blue curve is obtained from various inversions of the integral transform and constitutes a lower band of the total theoretical error-bar associated to this observable.
  A very good description of this quantity with respect to the experimental data from Ref.~\cite{ahrens1975} is obtained already at this level of approximation in coupled-cluster theory. This result highlights the relevance of the newly introduced method, which allowed $(i)$ to depart from the classical few-body systems and reach  medium mass nuclei  with only little more computational expense and $(ii)$ to obtain, for the first time, a microscopic description of experimental data in this mass range.

\section{The electric dipole polarizability}

We now focus our attention on the electric dipole polarizability. The latter is defined as 
 an inverse energy weighted sum rule of the dipole response function
\begin{equation}\label{polresp}
\alpha_D = 2\alpha \int_{\omega_{ex}}^{\infty}  d\omega~\frac{R(\omega)}{\omega}\,.
\end{equation}
It is well known from the nuclear droplet and hydrodynamic models~\cite{Lipparini89}, as well as from density functional theory~\cite{Roca-Maza2015},
that the electric dipole polarizability is correlated with the nuclear charge radius $r_{ch}$. Thus, it is interesting to see if this correlation is also observed in first principle calculations and the developed LIT-CC technology allows for such a study.

In order to study correlations, one needs a considerable number of interactions, such that  results span a wide range of values both for the polarizability and the charge radius. For this purpose, we take some of the few realistic nucleon-nucleon interaction and evolve them with similarity transformations. 
\begin{figure}
\begin{center}
  \includegraphics*[width=15cm]{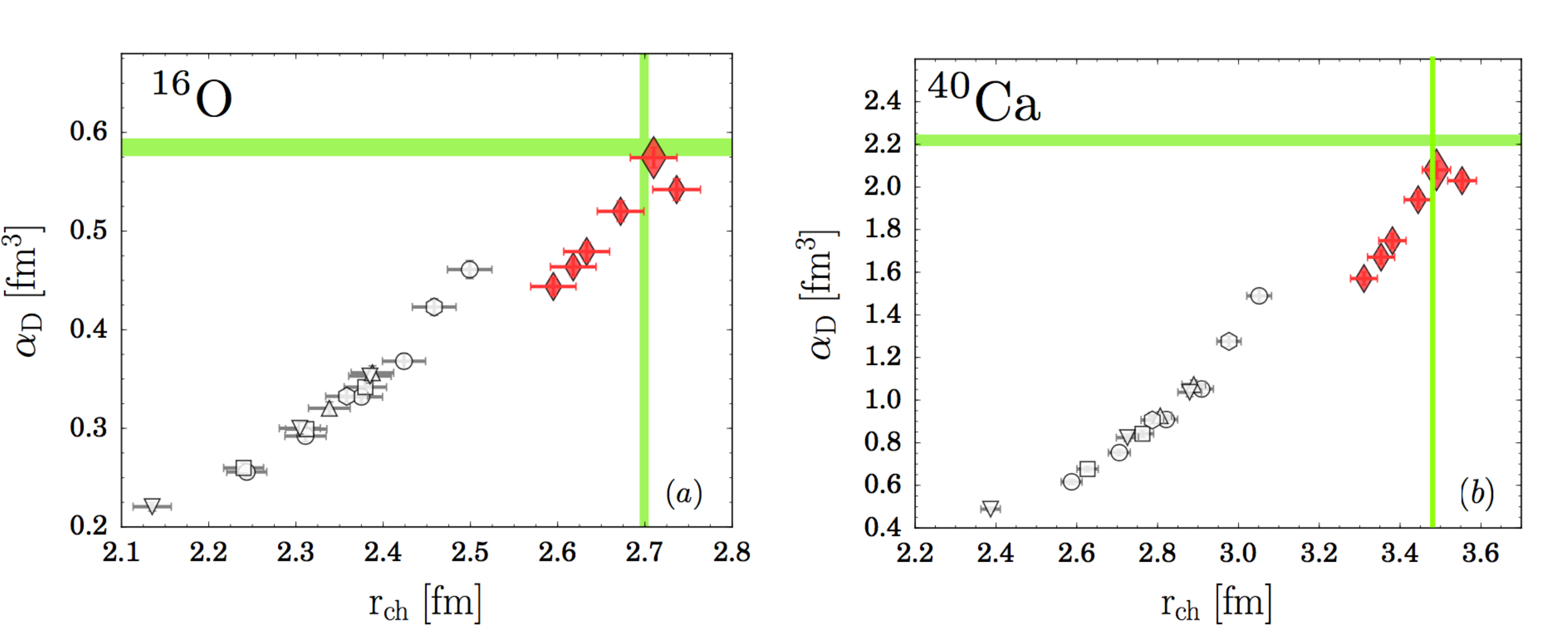}
\end{center}
\caption{ \label{fig_pol} $\alpha_D$ versus $r_{ch}$ in \textsuperscript{16}O (a) and \textsuperscript{40}Ca (b).
Empty symbols refer to calculations with nucleon-nucleon potentials only. The red (full) diamonds refer to calculations that include three-body forces. Details on the potentials used are found in Ref.~\cite{Miorelli}. The green bands denote the experimental values for these two quantities. Figure adapted from Ref.~\cite{Miorelli}.}
\end{figure}
Fig.~\ref{fig_pol} shows the correlation of the electric dipole polarizability $\alpha_D$ with the charge radius in $^{16}$O (a) and in $^{40}$Ca (b). The empty symbols refer to calculations with two-body forces only. They clearly show a strong correlation between the two observables. On the other hand, one also observes that the use of two-body interactions only leads to a strong underestimation of the experimental values of both quantities, denoted by the green bands. 
When we include three-nucleon forces, by using a variety of Hamiltonians from chiral effective field theory~\cite{Hebeler11,Ekstroem15}, we obtain the solid diamonds, which still exhibiting a strong correlation but cluster much closer to the experimental value. This fact indicates that three nucleon forces are important to obtain a correct description of both observables. It needs to be noted though, that we have made a specific choice of Hamiltonians, preferring those that provide a reasonable description of the charge radius. It is well known that other Hamiltonians from chiral effective field theory suffer from and under-estimation of the radii, which, in the light of Fig.~\ref{fig_pol}, would also lead to an underestimation of the electric dipole polarizability.
In particular, the largest diamond data are obtained with the  NNLO$_{sat}$~\cite{Ekstroem15} potential, which has been fit to reproduce also the $^{16}$O charge radius. It is interesting to note that this constraint enables a correct prediction also for $^{40}$Ca.

By studying similar correlations in  $^{48}$Ca and combining them with  experimental information from electron scattering, we also were able to provide narrow constraints on the neutron skin thickness and on the electric dipole polarizability~\cite{Hagen16,Birkhan17}.

\section{The Coulomb sum rule}
Finally, we turn our attention to the Coulomb sum rule.
With the objective in mind to eventually tackle neutrino-nucleus interactions relevant for long-baseline neutrino experiments,
we initiated a study of electron-scattering reactions within coupled-cluster theory.
 Following  the work done with the Green's function Monte Carlo method in Ref.~\cite{Lovato}, we  first investigate the Coulomb sum rule.

 Omitting for simplicity the role of the nucleon form factor, the longitudinal Coulomb sum rule  is defined as
 as total strength of the inelastic longitudinal response function as
\begin{equation} 
CSR(q)=\int_{\omega^+_{th}}^\infty d\omega R_L(\omega,q)\,,
\label{eq:lcsr}
\end{equation}
where
\begin{equation}
R_L(\omega,q)=\int \!\!\!\!\!\!\!\sum _{f\ne 0} 
|\left\langle \Psi_{f}| \hat\rho(q)| 
\Psi _{0}\right\rangle|
^{2}\delta\left(E_{f}+\frac{q^2}{2M}-E_{0}-\omega \right),\,
\end{equation}
with $\hat\rho(q)$ being the charge operator and
$\frac{\mathbf{q}^2}{2M}$ a recoil term.
One possible approach to compute the Coulomb sum rule  is to perform a multipole expansion of the charge operator and then solve Eq.~(\ref{psi1}) for every multipole on a grid of $q$, as was done in Ref.~\cite{th4he}. However, 
the Coulomb sum rule  can be rewritten also as expectation value on the ground state which is much easier to calculate
\begin{equation}
 CSR(q)=Z+ \langle \psi_0| \sum_{i\ne j} e^{i{\bf q}\cdot{( {\bf r}_i- {\bf r}_j)}}| \psi_0 \rangle - |F_{el}(q)|^2 Z^2 = Z+Z(Z-1)f_2(q) - |F_{el}(q)|^2 Z^2\,,
\end{equation}
where $f_2(q)$ is the Fourier transform of the proton-proton correlation density and $F_{el}(q)$ is the elastic form factor~\cite{Giusy}.
\begin{figure}
  \begin{center}
    \label{bench}
  \includegraphics*[width=15cm]{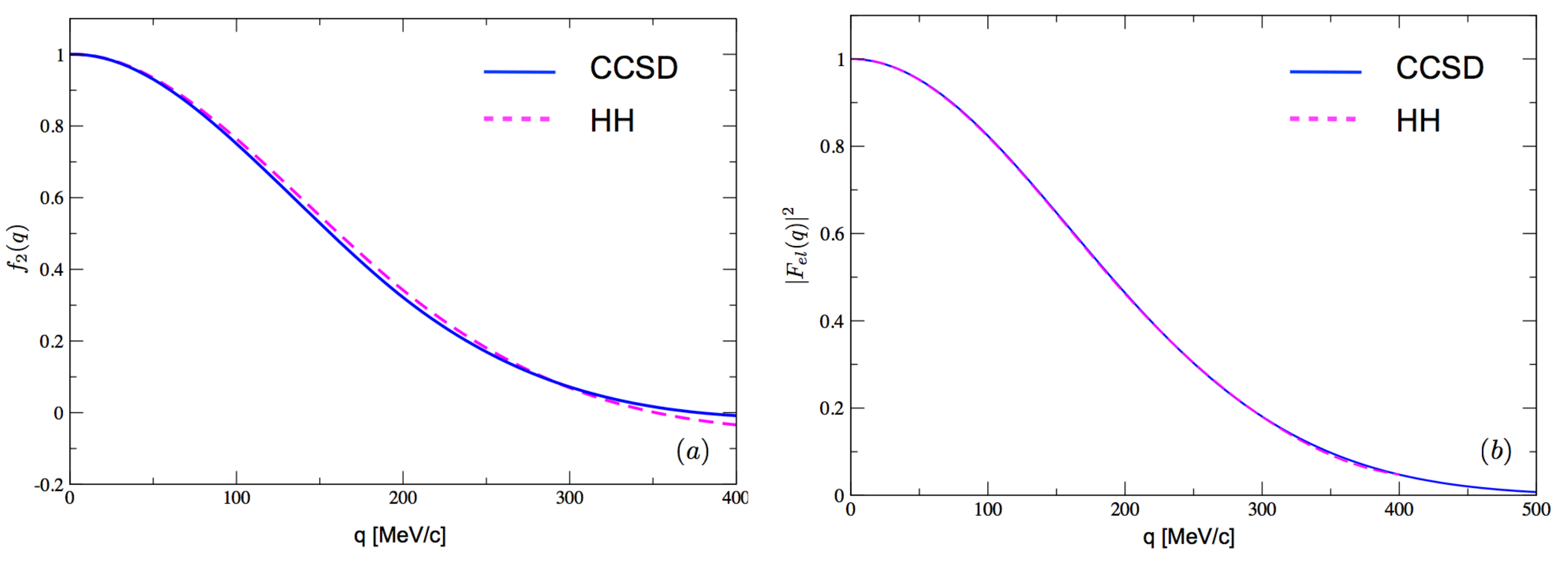}
  \end{center}
\caption{ \label{fig_csr} Comparison of CCSD and exact hyperspherical harmonics (HH) computations of $f_2(q)2$, panel (a) and $|F_{el}(q)|^2$, panel (b) as a function of the momentum transfer $q$ using a chiral two-body force~\cite{Entem03}. \\}
\end{figure}
To first compare the CCSD approximation with exact calculation we compare computations of $f_2(q)$ and $|F_{el}(q)|^2$ with exact hyperspherical harmonics, as shown in Fig.~\ref{bench} vs. the momentum transfer. Calculations are performed with a chiral two-body force~\cite{Entem03} and show a very nice agreement between the two methods, proving the CCSD approximation to be very reliable. 
\begin{figure}
\begin{center}
  \includegraphics*[width=15cm]{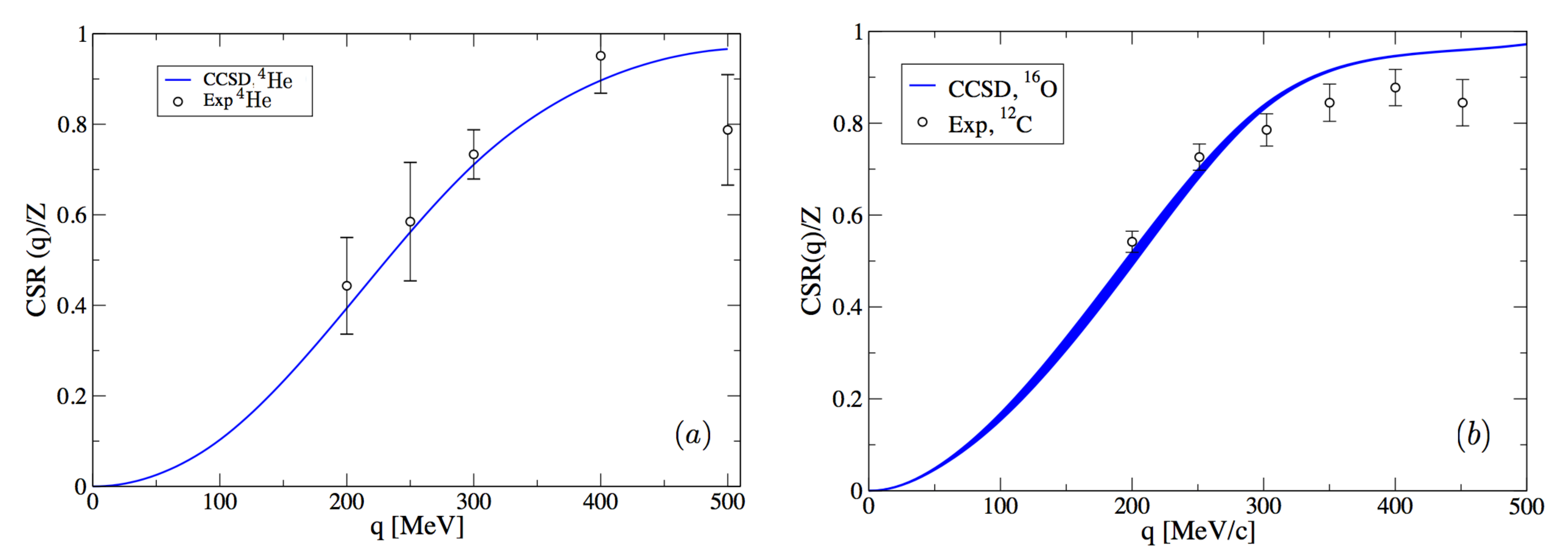}
  \end{center}
\caption{ \label{fig_csr}   Coulomb sum rule for $^4$He (a) and $^{16}$O (b) from coupled-cluster theory using a chiral two-body force~\cite{Entem03}, compared to experimental results  from \cite{buki,world_data} for $^4$He (a), and from Ref.~\cite{Mihaila} for  $^{12}$C (b).}
\end{figure}

In Fig.~\ref{fig_csr} we finally compare the  Coulomb sum rule  calculated from coupled-cluster theory for $^4$He and $^{16}$O with 
 experimental data.   Measurements of the longitudinal response functions at 
 low momentum transfers ($q=200$ and $250$ MeV/c) are taken from Buki {\it et al.}~\cite{buki}, while 
 intermediate momentum transfer values are collected in Ref.~\cite{world_data}. Unfortunately, no data exist for $^{16}$O, so we show data for $^{12}$C taken from~\cite{Mihaila}. Similar results were also obtained in Refs.~\cite{Mihaila,Lonardoni}. Fig.~\ref{fig_csr} shows that our calculations agree quite well with the experimental data. Thus, extensions to heavier nuclei and to neutrino reactions are promising.

\section{Conclusions}
\label{sec-concl}
We reviewed some of our recent results on electromagnetic reactions and related sum rules obtained from  coupled cluster theory and the Lorentz integral transform. The method has allowed us to surpass previous mass limits and paves the way for ab initio studies of neutrino-nucleus cross section. We addressed first electron scattering observables, such as the Coulomb sum rule, and performed a comparison to exact calculations and to data. This is essential for any theory that is going to be used to model  neutrino-nucleus interactions.

\subsection{Acknowledgments}
This work was supported in parts by the Natural Sciences and Engineering Research Council (NSERC), the National Research Council of Canada, the PRISMA Cluster of Excellence and the Sonderforschungbereich CRC 1044.


\section*{References}


\begin{thebibliography}{9}

\bibitem{GoT48}  Goldhaber M and Teller E 1948 Phys. Rev. {\bf 74} 1046

\bibitem{steinwedel1950} Steinwedel H and Jensen  J H D 1950  Z. Naturforsch  {\bf 5A} 413

\bibitem{Bacca:2013dma} Bacca S {\it et al.} 2013
   Phys. Rev. Lett. {\bf 111} 122502


   \bibitem{Efl07} Efros V D {\it et al} 2007
 Journal of Physics G: Nuclear and Particle Physics
 {\bf 34} R459

\bibitem{Epelbaum09} Epelbaum E,  Hammer H-W and Mei\ss{}ner U-G 2009
  Rev. Mod. Phys. {\bf 81} 1773

\bibitem{Machleidt11}  Machleidt R  and Entem D R  2011 
Physics Reports {\bf 503} 1
 
 \bibitem{barnea2001} Barnea N, Leidemann W and Orlandini G 2001  Nuclear Physics A {\bf 693}
565

\bibitem{hagen2013c} Hagen G, Papenbrock T, Hjorth-Jensen M and Dean D J 2014
 Rep. Prog. Phys. {\bf 77} 096302

  
\bibitem{coester1960} Coester F and K{\"u}mmel H 1960
Nucl. Phys. {\bf 17} 477


\bibitem{Mirko}  Bacca S, Barnea N, Hagen G, Miorelli M, Orlandini G and Papenbrock T 2014
 Phys. Rev. C {\bf 90} 064619

 \bibitem{Entem03}
Entem D R  and  Machleidt R 2003 Phys. Rev. C \textbf{68} 041001

 
\bibitem{ahrens1975}
Ahrens J {\it et al.} 1975
Nucl. Phys. A
{\bf 251}, 479



\bibitem{Lipparini89} Lipparini E and Stringari S 1989 Phys. Rep. {\bf 175} 104

\bibitem{Roca-Maza2015} Roca-Maza X {\it et al.} 2015
  Phys. Rev. C {\bf 82} 064304 

  
\bibitem{Miorelli} Miorelli M,  Bacca S,  Barnea N,  Hagen G,  Jansen G R, Orlandini G,  Papenbrock T 2016
Phys.~Rev.~C {\bf 94} 034317 



 
\bibitem{Hebeler11} Hebeler K  and Bogner S K,  Furnstahl R J,  Nogga A  and Schwenk  A 2011 Phys. Rev. C {\bf 83},
  031301

\bibitem{Ekstroem15} Ekstr\"om A 
  {\it et al.} 2015
 Phys. Rev. C, {\bf 91} 051301

  

  
\bibitem{Hagen16}
   Hagen G  {\it et al.} 2016
 Nature Physics {\bf 12}, 186

\bibitem{Birkhan17}
  Birkhan J {\it et al.} 2017
 Phys. Rev. Lett. {\bf 118} 252501



\bibitem{Lovato} A.~Lovato {\it et al.} 2013
  Phys.~Rev.~Lett. {\bf 111} 092501

\bibitem{th4he}
 Bacca S,  Barnea N,  Leidemann W and  Orlandini G 2009 Phys. Rev. C \textbf{80} 064001 


\bibitem{Giusy} Orlandini G and Traini M 1991 Rep. Prog. Phys.~{\bf 54} 257

\bibitem{buki}  Buki A {\it et al.} 2006
  Phys. Lett. B \textbf{641}, 156 

\bibitem{world_data}  Carlson J,  Jourdan J,  Schiavilla R  and  Sick I 2002 Phys. Rev. C
{\bf 65}, 024002 
\bibitem{Mihaila}  Mihaila B and Heisenberg H J 2000  Phys. Rev. Lett. {\bf 84} 1403
\bibitem{Lonardoni} Lonardoni D {\it et al. } 2017  Phys. Rev. C {\bf 96} 024326

\end{thebibliography}
\end{document}